# Theoretical Model of Acoustic Power Transfer Through Solids

Ippokratis Kochliaridis and Michail E. Kiziroglou,

Department of Industrial Engineering and Management, International Hellenic University, Greece
**Email:** m.kiziroglou@gmail.com , Ippokratis_polichni@hotmail.com

*Abstract*—Acoustic Power Transfer is a relatively new technology. It i's a modern type of a wireless interface, where data signals and supply voltages are transmitted, with the use of Mechanical Waves, through a medium [1][2]. The simplest application of such systems is the measurement of frequency response for audio speakers. It consists of a variable Signal Generator, a Measuring Amplifier which drives an acoustic source and the Loudspeaker Driver. The receiver contains a Microphone Circuit with a Level Recorder [3]. Acoustic Power Transfer could have many applications, such as: Cochlear Implants, Sonar Systems and Wireless Charging. However, it's a new technology, so it needs further investigation.

## I. Introduction

There are many wireless interfaces, such as: RFID. However, these technologies in some cases, could be problematic. For example, electromagnetic signals couldn't penetrate metallic surfaces, due to Faraday Shielding Effect, or are limited to a very low Bandwidth (<100 Hz) [4][5]. In addition, there are many marine sensor applications, which require penetration thought solid metal walls. In such cases, drilling is not preferable as a solution for signal or power coupling through metallic walls, due to reliability and sealing problems. As a result, data and power supply transmission using acoustic waves, from an acoustic source to a receiver, could be a promising solution [2].

## II. System Arrangement

The figure 1 represents the Block Diagram, of a simplified Acoustic Power Transfer System. The Transmitter (TX) consists of a signal generator, a power amplifier and an acoustic source, which is a Piezoelectric Ultrasound Driver [1]. The Receiver consist of a Piezoelectric Microphone and a network circuit, which charges a battery. The transmitter uses the piezoelectric effect to generate acoustic waves from electrical power [1]. The receiver converts acoustic waves into electrical power, again using the piezoelectric effect. The medium, could be air, metals and even tissue in the case of medical implants.

### A. An application system

Figure 2 shows a typical arrangement of an Acoustic Power Transfer System. It consist of a medium of transfer , which is a metallic surface, an acoustic source, which is a piezoelectric ultrasound driver and the receiver in the other side which is a piezoelectric transducer also. The gap between the metallic wall and the diaphragms is filled with some type of glue suitable for acoustic interfacing. This arrangement is often referred to as the -One Dimension Propagative Model (ODPM) [2].

### B. Characteristics of the System

A study has shown that this type of arrangement could transmit power over 100 W, for 10 min and with 60% efficiency, though a 57.2 mm steel surface [6]. Lawry et al [7] have proposed a method to calculate load impedance, for the receiver output, in order to achieve higher efficiency.

Beyond Wireless Power Transfer, this arrangement can also be used for acoustic data transfer. However the Data Rate is relatively small, in the range of 55 kbps [8]. In order to achieve higher data rates, signal processing technics, such as Orthogonal Frequency Division Multiplexing (OFDM) [2] should to be used. In a case of an Array, with seven Piezoelectric Drivers and Microphones, in the Transmitter and the Receiver respectively, a maximum Data Rate of up to 700 Mbps was demonstrated [9].

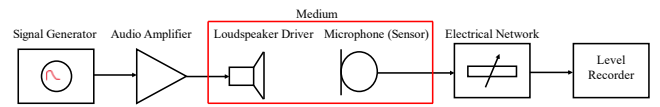

Fig. 1. A simplified block diagram, which illustrates the transmitter and the receiver.

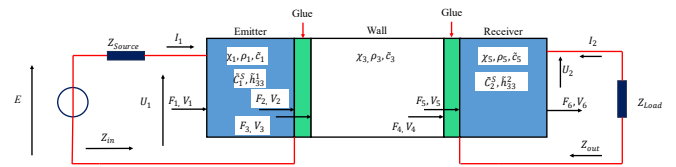

Fig. 2. An arrangement of an Acoustic Power of Transfer System according to Freychet et al [2].

## C. Modeling the System – Equivalent Circuits

The technical characteristics of a Piezoelectric Driver or a Receiver, could be expressed as an equivalent electrical circuit network. The most common mathematical models are the Mason model and the Krimholtz, Leedom, and Matthaei (KLM) [10]. The Mason concept was to present Electrical, Mechanical and Acoustical properties of a transducer, as an electrical line. The Figure 3 represents the mason equivalent electrical system.

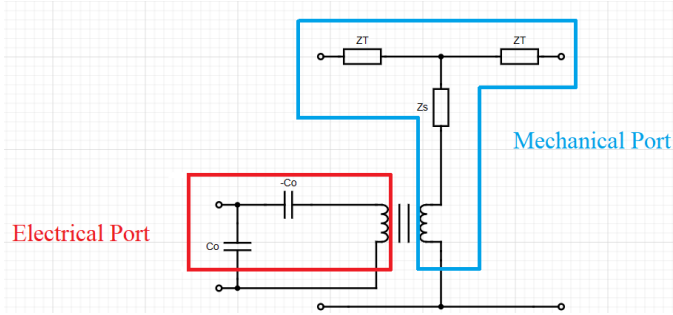

Fig. 3 The equivalent electrical network of the mason model.

Figure 3 illustrates that a piezoelectric transducer, consists of a mechanical part (the network with 3 blue Impedances) and an electrical stage (the positive and negative capacity). The KLM model is simpler, because it eliminates the negative capacity [10]. The figure 4 illustrates the mathematical concept of Mason, applied in a complete electro-acoustical signal transfer system:

The first stage (Green), illustrates the equivalent electrical circuit of the Acoustic Source. The stage with the 3 blue Impedances, represents the Matching Layer. This corresponds to the gap between the diaphragm of the driver and the metallic surface. The Orange stage represents the impedance of the metallic medium. Finally, the red impedances is the receiver circuit.

The complete set of equations which describe the Acoustic Power Transfer System, given in figure 2, could be summarized into a single matrix given in equation (1) [2]:

$$\begin{bmatrix} F_1 \\ -F_2 \\ U_1 \\ F_2 \\ -F_3 \\ F_3 \\ -F_4 \\ F_4 \\ -F_5 \\ F_5 \\ -F_6 \\ U_2 \end{bmatrix} = \begin{bmatrix} X_{11}^1 & X_{12}^1 & X_{13}^1 & 0 & 0 & 0 & 0 & 0 \\ X_{12}^1 & X_{11}^1 & -X_{13}^1 & 0 & 0 & 0 & 0 & 0 \\ X_{13}^1 & -X_{13}^1 & X_{33}^1 & 0 & 0 & 0 & 0 & 0 \\ 0 & X_{11}^2 & 0 & X_{12}^2 & 0 & 0 & 0 & 0 \\ 0 & X_{12}^2 & 0 & X_{11}^2 & 0 & 0 & 0 & 0 \\ 0 & 0 & 0 & X_{11}^3 & X_{12}^3 & 0 & 0 & 0 \\ 0 & 0 & 0 & X_{12}^3 & X_{11}^3 & 0 & 0 & 0 \\ 0 & 0 & 0 & 0 & X_{11}^4 & X_{12}^4 & 0 & 0 \\ 0 & 0 & 0 & 0 & X_{12}^4 & X_{11}^4 & 0 & 0 \\ 0 & 0 & 0 & 0 & 0 & X_{11}^5 & X_{12}^5 & X_{13}^5 \\ 0 & 0 & 0 & 0 & 0 & X_{12}^5 & X_{11}^5 & -X_{13}^5 \\ 0 & 0 & 0 & 0 & 0 & X_{13}^5 & -X_{13}^5 & X_{33}^5 \end{bmatrix} \begin{bmatrix} V_1 \\ V_2 \\ I_1 \\ V_3 \\ V_4 \\ V_5 \\ V_6 \\ I_2 \end{bmatrix}$$

(5)

In this matrix for a five layer arrangement, $V_n$ corresponds to the speed at the nth interface, and $F_n$ corresponds to the force at the nth interface acting on the material at the right of the interface [2]. $U_n$ are the voltages, which apply across two consecutive layers of the transducers, $I_n$ are the currents and finally $X_n$ are related to the Acoustic Characteristic Impendance of each layer. An example which demonstrates the practical use of this Transfer-Function Matrix, can be found in the determination of Input Force $F_1$ (It's the force which applies, in the first side of the Transducer-Driver) as follows:

$$F_1 = X_{11}^1 * V_1 + X_{12}^1 * V_2 + X_{13}^1 * I_1 + 0*V_3 + 0*V_4 + 0*V_5 + \ldots 0*I_2 \quad (1).$$

where: $V_1$ is the velocity which applies in the first side of the Transducer-Driver, $X_{11}^1$ is the Acoustic Impedance of the transducer in the first side, $V_2$ is the velocity which applies in the second side of the Transducer, $X_{12}^1$ is accordingly the Acoustic Impendence in the second side of the Transducer, $X_{13}^1$ is the Acoustic Impendence of whole Transmitter and $I_1$ is the supply current of the Acoustic Source.

The 1st matrix's element (Acoustic Impendence $X_{11}^1$), according to Freychet [2], is a result from the mathematical relationship 2:

$$X_{11}^n = \frac{\tilde{Z}_n^m}{\tan(\tilde{k}_n x_n)} \quad (2).$$

Where: $x_n$ is the surface (layer) thickness. $\tilde{k}_n$ is the wave number [2] given by equation 3:

$$\tilde{k}_n = \frac{\omega}{\tilde{C}_n} \quad (3),$$

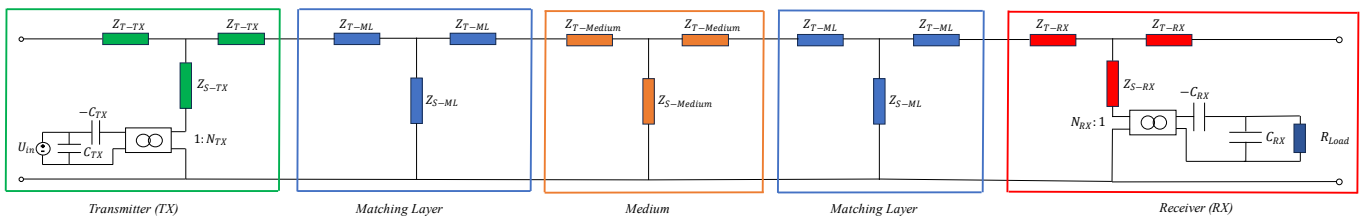

Fig. 4. The equivalent electrical network in the whole System.

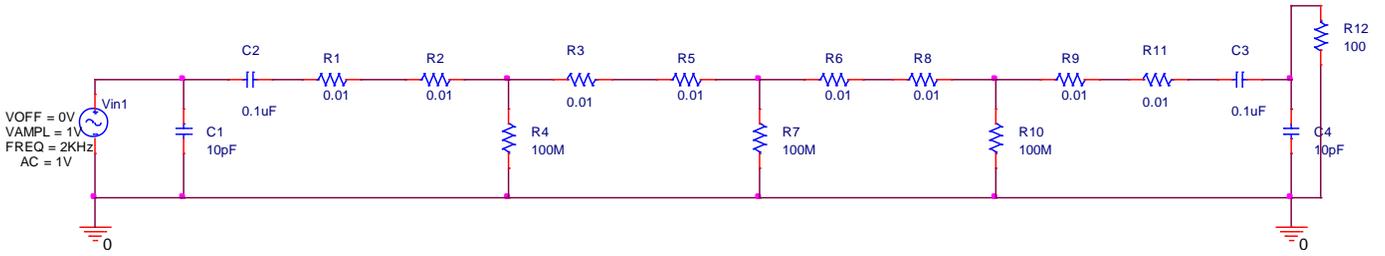

Fig. 6 A simplified electrical network of Mason's equivalent circuit, for future investigation.

Here, ω is the signal angular frequency. Finally, $\tilde{Z}_n^m$, is the Mechanical Characteristic Impendence of the layer [2], which is defined by equation 4:

$$\tilde{Z}_n^m = S\tilde{Z}_n^a = S\rho_n \tilde{c}_n \quad (4).$$

In this equation $\rho_n$ is the density, S the surface of the layer and $\tilde{c}_n$ the longitudinal speed of sound.

### III. Conclusion

In this paper a theoretical analysis of acoustic power transfer through solids was presented. The analysis is based on the standard models of Freychet and Mason. A corresponding simulation model was designed using the Cadence circuit design software, as indicatively shown in Figure 6. Subsequent steps of this work involves calculation of acoustic power transfer through metal plates and experimental work to investigate the possibility of using acoustic lenses and reflectors in order to maximize power deliver to specific locations.